# Collisional mixing between inner and outer solar system planetesimals inferred from the Nedagolla iron meteorite

Fridolin Spitzer[1], Christoph Burkhardt[1], Jonas Pape[1], and Thorsten Kleine[1]

[1]Institut für Planetologie, University of Münster, Wilhelm-Klemm-Str. 10, 48149 Münster, Germany

Corresponding author: Fridolin Spitzer (fridolin.spitzer@uni-muenster.de)

Christoph Burkhardt (burkhardt@uni-muenster.de)

Jonas Pape (jonas.pape@uni-muenster.de)

Thorsten Kleine (thorsten.kleine@uni-muenster.de)

# Abstract

The ungrouped iron meteorite Nedagolla is the first meteorite with bulk Mo, Ru, and Ni isotopic compositions that are intermediate between those of the non-carbonaceous (NC) and carbonaceous (CC) meteorite reservoirs. The Hf-W chronology of Nedagolla indicates that this mixed NC-CC isotopic composition was established relatively late, more than 7 million years after Solar System formation. The mixed NC-CC isotopic composition is consistent with the chemical composition of Nedagolla, which combines signatures of metal segregation under more oxidizing conditions (relative depletions in Mo and W), characteristic for CC bodies, and more reducing conditions (high Si and Cr contents), characteristic for some NC bodies, in a single sample. These data combined suggest that Nedagolla formed as the result of collisional mixing of NC and CC core material, which partially re-equilibrated with silicate mantle material that predominantly derives from the NC body. These mixing processes might have occurred during a hit-and-run collision between two differentiated bodies, which also provides a possible pathway for Nedagolla's extreme volatile element depletion. As such, Nedagolla provides the first isotopic evidence for early collisional mixing of NC and CC bodies that is expected as a result of Jupiter's growth.

# Introduction

The fundamental isotopic dichotomy between *non-carbonaceous* (NC) and *carbonaceous* (CC) meteorites (Budde et al. 2016a; Warren 2011) indicates that their parent bodies formed in two distinct areas of the solar accretion disk that remained spatially separated for several million years (Ma) (Kruijer et al. 2017). It has been suggested that the formation of Jupiter is responsible for the initial separation of the NC and CC reservoirs, where the NC reservoir represents the inner disk, whereas the CC reservoir represents the outer disk [for recent reviews see Bermingham et al. (2020), Kleine et al. (2020), and Kruijer et al. (2020)]. However, NC and CC meteorites both derive from parent asteroids located in the present-day asteroid belt between Mars and Jupiter, suggesting that at some point after their formation, CC-type bodies were scattered into the inner solar system and implanted into the asteroid belt. This is consistent with the expected gravitational effects of Jupiter's growth and migration on the planetesimal population inside and outside its orbit (Raymond and Izidoro 2017; Walsh et al. 2011). The inward scattering of CC bodies likely led to collisions among NC- and CC-type bodies, which may have produced mixed NC-CC isotopic signatures. Although there is evidence for collisional mixing of NC and CC materials in the form of CC clasts in brecciated NC meteorites (e.g., Goodrich et al. 2021; Patzek et al. 2018), until now no mixed NC-CC isotopic signature has been observed in bulk meteorites. Identifying such a signature would be of considerable interest, as it would provide direct empirical evidence for mixing of NC- and CC-type materials, and may allow constraining the timescale of this mixing.

Here we report bulk chemical, Ni, Mo, Ru, and Pt isotopic, and Hf-W chronological data for the ungrouped iron meteorite Nedagolla. These data are used to show that Nedagolla is the first bulk meteorite for which a mixed NC-CC isotopic composition is identified. The chemical and isotopic data are then used to assess the nature and origin of the NC and CC bodies involved in the mixing that produced Nedagolla, while the Hf-W data are used to constrain the timescale of mixing and metal segregation on the Nedagolla parent body.

# Materials and Methods

## Nedagolla

The ungrouped iron meteorite Nedagolla is an observed fall from 1870 in India with a total recovered mass of ~4.5 kg (Buchwald 1975). It is an anomalous nickel-poor ataxite and is characterized by an unusual chemical composition, including the lowest Ge content (Schaudy et al. 1972) (Fig. 1) and one of the highest Si and Cr concentrations observed among iron meteorites (Wai and Wasson 1970). The latter indicate formation under stronger reducing conditions than most other iron meteorites, because under more oxidizing conditions these elements are preferentially retained in the silicate compared to metal (Wai and Wasson 1970). Formation under more reducing conditions is also consistent with the presence of graphite between the metallic dendrites (Buchwald, 1975). The dendritic structure of Nedagolla implies rapid cooling within a few days (0.02°C/sec), most likely as a result of impact melting near the surface of its parent body (Miyake and Goldstein, 1974). In addition to graphite, Nedagolla contains daubréelite in the interdendritic pockets, but silicates and iron sulfide appear to be absent. The absence of iron sulfide in particular is rare among impact-melted iron meteorites, which are typically S-rich (e.g., Scott 2020). More recently, Walker et al. (2005) showed that Nedagolla is characterized by broadly chondritic relative abundances of Re, Os, Ir, Ru, and Pt, but is depleted in the less refractory Pd, which is even lower than in the strongly volatile-depleted IVB irons (Fig. 2).

## Analytical methods

The sample of this study was obtained from the National History Museum London (item number BM.1985,M268) as part of a larger systematic study on the Mo and W isotopic compositions of a total of 26 ungrouped iron meteorites (Spitzer et al. 2020a). Owing to the unusual Mo isotopic composition of Nedagolla, this meteorite was then further investigated for its Ni and Ru isotopic compositions. The results for Nedagolla are reported in this study, while the results for the other ungrouped iron meteorites will be published separately.

Three individual pieces (UI-05a, UI-05b, UI-05c) of 560 mg, 492 mg, and 279 mg, respectively, were cut from the allocated mass of Nedagolla using a diamond saw, polished with abrasives (SiC) and subsequently ultrasonicated in ethanol to remove any saw marks and adhering dust. The three subsamples were digested in double distilled 6M HCl (+trace $cHNO_3$) on a hot plate at 130 °C for at least 24 h. Upon complete dissolution, an aliquot representing 50 mg of material was taken from UI-05a and UI-05b for Pt isotope analyses, while the remainder was processed for W and Mo isotope analyses. For UI-05a and UI-05c an additional aliquot corresponding to 10 mg and 56 mg, respectively, was taken for Ni isotope analyses, while the remainder of UI-05c was processed for Ru isotope analysis. The bulk chemical composition of a solution aliquot of UI-05a was measured on a Thermo Scientific X-Series quadrupole ICP-MS.

The chemical separation and purification of W, Mo, Pt, Ni, and Ru for isotope measurements followed our previously established protocols and are only briefly summarized here. The separation of W and Mo followed the analytical protocols described in Budde et al. (2018) and Kruijer et al. (2017), where the Mo cuts collected during both steps of the W purification were subsequently purified using a two-stage ion exchange chromatography slightly modified from Burkhardt et al. (2011, 2014). The yields for W and Mo were ~60% and ~75%, respectively. The chemical separation of Pt followed the protocol described in Kruijer et al. (2013), which is based on techniques initially developed by Rehkämper and Halliday (1997) and uses a single-stage anion exchange column for the separation of Pt. The total yields of this procedure were ~ 80%. The chemical purification of Ni involved a three-step ion exchange chromatographic procedure following the protocols described in Nanne et al. (2019), which are based on the method of Chernonozhkin et al. (2015). The total yield for Ni was ~80%. Finally, the Ru purification procedure largely followed established protocols from this laboratory (Fischer-Gödde et al. 2015; Hopp and Kleine 2018), but were slightly adjusted owing to the unusual chemistry of Nedagolla. After complete digestion in double-distilled 6M HCl, the sample solution was converted to reverse aqua regia by adding concentrated $HNO_3$ and equilibrated overnight on a hot plate at 120°C. The sample solution was then dried down at temperatures below 80°C to near dryness, taken up and dried down twice in 6M HCl and once in 1M HCl at slightly elevated temperatures. The sample was finally dissolved in 0.2M HCl and loaded onto cation exchange columns loaded with 10ml of BioRad AG 50 W-X8 (100-200 mesh). During this column chromatography step Ru together with other HSEs is separated from major and minor matrix elements like Fe, Ni and Cr. Due to Nedagolla's elevated Cr concentrations this step was repeated four times in order to remove left-over Cr from the sample matrix and to achieve a clean-enough sample before subsequent Ru purification by microdistillation. Typical Ru yields of the cation columns were >95%. Ruthenium was finally purified by microdistillation (Fischer-Gödde et al. 2015), where the sample was loaded into the cap of a convex 5ml Savillex beaker and dissolved in a $H_2SO_4$-$CrO_3$ solution. During stepwise heating, Ru was oxidized and evaporated into a droplet of concentrated HBr. After microdistillation the HBr containing the purified Ru fraction was dried down and finally taken up twice in 0.28M $HNO_3$ for isotope analysis. The Ru yield of the microdistillation was about 37%. Although this is on the lower side of what is typically achieved during microdistillation (~30-80%), several experiments demonstrate that such low distillation yields do not impact upon the accuracy of $\varepsilon^{100}$Ru data (Bermingham et al. 2016). Total procedural blanks for W, Mo, Pt, Ni, and Ru were all negligible.

All isotope measurements were performed on a Thermo Scientific Neptune *Plus* MC-ICP-MS at the Institut für Planetologie in Münster, using the established measurement protocols described in the aforementioned studies from this laboratory. All isotope data are reported in ε-notation (i.e., parts-per-10,000 deviations from the mean values of terrestrial solution standards that were analyzed bracketing the sample measurements). For samples analyzed several times, reported ε-values represent the mean of pooled solution replicates, and uncertainties are reported as 95% confidence intervals (CI) of the mean for a two-sided Student's *t*-distribution with *n*-1 degrees of freedom.

The accuracy and precision of the isotope measurements were assessed by repeated analyses of the NIST 129c and NIST 361 metal standards, which for Pt and Ru were doped with appropriate amounts of the terrestrial solution standard to match the concentrations of the analyzed samples. The external reproducibility (2 s.d.) of the isotope measurements of the NIST 129c and NIST 361 standards can be obtained from the data in Tables 1-5. They return isotopic compositions as reported in previous studies testifying to the accuracy of the isotope measurements. Tungsten isotope ratios involving $^{183}$W display a small mass-independent effect, which has also been observed in several previous high-precision W isotope studies (*e.g.*, Kruijer et al. 2012, Willbold et al. 2011). This effect results in small excesses for $\varepsilon^{182}$W (‘6/3’) and $\varepsilon^{184}$W (‘6/3’) and corresponding deficits in $\varepsilon^{183}$W (‘6/4’). The magnitude of this analytical effect varies between different studies and is typically about -0.1 to -0.2 for $\varepsilon^{183}$W and is likely induced during incomplete dissolution of chemically purified W in Savillex beakers. For the samples of this study, W isotope ratios involving $^{183}$W were, therefore, corrected using the mean $\varepsilon^{i}$W values obtained for the NIST 129c analyses of this study, using the method described in Kruijer et al. (2014). The associated uncertainties induced through this correction were propagated into the final reported uncertainties of the W isotope ratios.

# Results

## Chemical composition

The concentrations of selected siderophile elements obtained by solution quadrupole ICP-MS in this study are reported in Table 6 along with literature data. The new concentration data generally agree with prior results to within ±10%. As noted in Walker et al. (2005), the HSEs Os, Re, Ir, Ru, and Pt occur in approximately chondritic relative abundances and are enriched by a factor of ~10-12 relative to CI chondrites (Fig. 2). The more volatile HSEs Pd and Au are depleted relative to the refractory HSEs, which is consistent with the overall volatile-depleted chemical composition of Nedagolla (Fig. 2). Of note, the refractory siderophile elements W and Mo are depleted relative to the refractory HSE by a factor of ~2 and ~4, respectively. Finally, our results confirm the high Cr content of Nedagolla of ~2600 µg/g reported by Wai and Wasson (1970), which is the highest so far reported value among iron meteorites.

## Pt isotopes

The purpose of the Pt isotope measurements is the quantification of potential secondary neutron capture effects induced during cosmic ray exposure (CRE). Such CRE-effects have been shown to modify the isotope compositions of iron meteorites for Mo (e.g., Spitzer et al. 2020b; Worsham et al. 2017), Ru (e.g., Fischer-Gödde et al. 2015), and W (Kruijer et al. 2013; Wittig et al. 2013). Some of these prior studies have also shown that for these elements the pre-exposure isotope composition (i.e., unaffected by CRE) of a sample can be determined using Pt isotopes as the neutron dosimeter. For the major groups of iron meteorites this correction is typically done using the empirical correlation of the isotope ratio of interest with $\varepsilon^{196}$Pt, such that the pre-exposure composition is given by the intercept of the correlation line at $\varepsilon^{196}$Pt=0. However, for individual, ungrouped samples CRE-effects cannot be corrected in this manner, but the pre-exposure isotope composition can instead be calculated using the measured $\varepsilon^{196}$Pt of a sample and the empirical slope of the correlation line between the isotope ratio of interest and $\varepsilon^{196}$Pt defined by the major iron meteorite groups (Fischer-Gödde et al. 2015; Spitzer et al. 2020b). Note that $\varepsilon^{196}$Pt is used for the CRE-correction because for $\varepsilon^{192}$Pt and $\varepsilon^{194}$Pt the magnitude of the CRE-effects depends on a sample's Ir/Pt ratio (Kruijer et al. 2013).

The Pt isotopic data for Nedagolla are reported in Table 1. The results for the two different digestions agree within their respective uncertainties, and the mean composition of the two samples is used for the CRE-correction. The slightly positive $\varepsilon^{192}$Pt, $\varepsilon^{194}$Pt, and $\varepsilon^{196}$Pt values indicate the presence of CRE-effects, where the $\varepsilon^{196}$Pt value of ~0.16 requires significant correction of CRE-effects on the W, Ru, and Mo isotope composition of Nedagolla (see below).

## Mo, Ru, and Ni isotopes

The Mo, Ru, and Ni isotopic data for Nedagolla are provided in Tables 2-4. The CRE-corrected $\varepsilon^{i}$Mo values fall within the range of Mo isotope anomalies typically observed for NC meteorites. However, unlike all previously analyzed bulk meteorites, Nedagolla plots between the NC- and CC-lines in the $\varepsilon^{95}$Mo–$\varepsilon^{94}$Mo diagram (Fig. 3), making it the first identified bulk meteorite having a mixed NC-CC Mo isotopic composition.

The CRE-corrected $\varepsilon^{100}$Ru of Nedagolla of -0.17±0.08 is intermediate between the composition of enstatite and ordinary chondrites, and among the iron meteorites intermediate between IAB and IVA irons. By contrast, Nedagolla's Ru isotopic composition is distinct from most CC meteorites, which typically display $\varepsilon^{100}$Ru values around -1, where the only exception are CI chondrites (-0.24±0.13; Fischer-Gödde and Kleine 2017).

Unlike Mo and Ru, Ni isotopes are not significantly affected by CRE-effects (Cook et al. 2020), and so no correction is necessary for measured Ni isotope ratios. In a plot of $\varepsilon^{64}$Ni versus $\varepsilon^{62}$Ni (Fig. 4a), all meteorites plot along a single correlation line, where CC meteorites are characterized by positive $\varepsilon^{64}$Ni and $\varepsilon^{62}$Ni, and NC meteorites by negative $\varepsilon^{64}$Ni and $\varepsilon^{62}$Ni values (Fig. 4a) (Nanne et al. 2019; Steele et al. 2011). Nedagolla plots at the lower end of the CC field and its Ni isotopic composition overlaps with that of group IIF and IVB irons, which both belong to the CC group of meteorites. However, when also considering $\varepsilon^{60}$Ni (Fig. 4b), Nedagolla plots outside the CC field and, instead, seems to plot on the extension of the array defined by NC meteorites, Earth, and Mars. The only CC meteorite with a similar $\varepsilon^{60}$Ni value as Nedagolla are the CI chondrites, which, however, have more elevated $\varepsilon^{62}$Ni and $\varepsilon^{64}$Ni (Fig. 4a).

### W isotopes

The W isotopic data for Nedagolla are provided in Table 5. The disparate $\varepsilon^{182}$W (6/4) and $\varepsilon^{182}$W (6/3) values together with the slightly positive $\varepsilon^{183}$W value of 0.09±0.09 are consistent with the presence of small nucleosynthetic W isotope anomalies, which have previously been observed for CC irons (Kruijer et al. 2017; Qin et al. 2008; Worsham et al. 2019). Nucleosynthetic effects on $\varepsilon^{182}$W (6/3) are negligible, but are significant for $\varepsilon^{182}$W (6/4) and can be corrected using the empirical correlation between nucleosynthetic $\varepsilon^{182}$W (6/4) and $\varepsilon^{183}$W (6/4) variations observed in acid leachates of primitive chondrites (Burkhardt and Schönbächler 2015; Burkhardt et al. 2012), chondrules and matrix (Budde et al. 2016b), or Ca-Al-rich inclusions (Burkhardt et al. 2008; Kruijer et al. 2014a). After this correction the $\varepsilon^{182}$W (6/4) and $\varepsilon^{182}$W (6/3) values of Nedagolla agree within the analytical uncertainty of the W isotope measurements (Table 5). These $\varepsilon^{182}$W values must then also be corrected for CRE-effects, using the empirical $\varepsilon^{182}$W versus $\varepsilon^{196}$Pt slope (-1.320 ± 0.055) obtained from the major iron meteorite groups (Kruijer et al. 2017). The pre-exposure $\varepsilon^{182}$W (6/3) value of Nedagolla calculated in this manner is -2.81±0.12. This value is more radiogenic than pre-exposure $\varepsilon^{182}$W values of all magmatic iron meteorite groups (Fig. 5), but overlaps with pre-exposure $\varepsilon^{182}$W values observed for some non-magmatic IAB (e.g., Worsham et al. 2017) and IIE irons (Kruijer and Kleine 2019).

## Discussion

Some of the key features of Nedagolla determined in this study may be summarized as follows. Nedagolla has a mixed NC-CC Mo isotopic composition, as is evident from its position between the NC- and CC-lines in the $\varepsilon^{95}$Mo–$\varepsilon^{94}$Mo diagram (Fig. 3). Nedagolla also has a peculiar Ni isotope composition, because it plots in the CC field in the $\varepsilon^{64}$Ni–$\varepsilon^{62}$Ni diagram, but on the extension of the NC field in the $\varepsilon^{60}$Ni–$\varepsilon^{62}$Ni diagram (Fig. 4b). By contrast, Nedagolla's Ru isotopic composition is similar to NC meteorites, such as enstatite and ordinary chondrites, and is distinct from most CC meteorites, with the exception of CI chondrites. Finally, Nedagolla has one of the most radiogenic $^{182}$W isotope compositions observed for iron meteorites, indicating a later metal-silicate fractionation than in most other iron meteorite parent bodies. These observations combined suggest that Nedagolla's formation involved mixing between NC and CC materials. This mixing may have involved some re-equilibration between metal and silicate, to account for Nedagolla's more radiogenic $^{182}$W composition compared to other magmatic iron meteorites. Below we will first discuss how Nedagolla's chemical composition also points towards a mixed NC-CC composition, then use the Mo-Ru-Ni isotopic data to constrain the nature and origin of the NC and CC materials that were mixed to produce Nedagolla, and use its $^{182}W$ composition to evaluate the timing of this mixing.

## Chemical constraints on formation conditions

The low S content of Nedagolla, which is evident from the absence of troilite, may either reflect complete exclusion of S-bearing molten metal during fractional crystallization or extreme volatile depletion during magmatic degassing. As such, the low S content suggests that Nedagolla was formed from differentiated material because complete removal of S-bearing molten metal requires fractional crystallization in a very large metal melt pool or core, and because only fully differentiated material from a core could be efficiently devolatilized by impact. This combined with the strong depletions in other volatile elements like Ge, whose abundances are, unlike S, not strongly affected by fractional crystallization, suggests that Nedagolla (or its precursor) derives from strongly devolatilized core material.

Other notable chemical features of Nedagolla include the depletions of W and Mo relative to other refractory siderophile elements (Fig. 2) and its high Cr and Si concentration, which are among the highest observed for iron meteorites. The presence of both signatures in a single iron meteorite is surprising, because they are indicative of oxidizing and reducing conditions, respectively, at the same time. Depletions of Mo and W relative to other refractory siderophile elements were previously observed for some CC irons, such as group IIC, IIF, and IVB irons, and are commonly interpreted as evidence for more oxidizing conditions during metal-silicate fractionation, which results in a less siderophile character of W and Mo (e.g., Campbell and Humayun 2005; Tornabene et al. 2020). More oxidizing conditions during metal segregation are also consistent with the comparatively high concentrations of most HSE (e.g., Os, Re, Ir, Ru), which argue for a proportionally smaller core and, therefore, more oxidized body. However, the metal-silicate partitioning of Cr is also strongly affected by the redox conditions, and the concentration of Cr in the metal decreases towards more oxidizing conditions (e.g., Wood et al. 2008). Thus, the high Cr concentration of Nedagolla (i.e., high compared to other iron meteorites) indicates formation under stronger reducing conditions than in other iron meteorites. This is also consistent with the elevated Si (e.g., Wai and Wasson 1970) and C concentrations of Nedagolla (Buchwald 1975). Nedagolla, therefore, displays chemical signatures for metal segregation under both more oxidizing conditions (i.e., depletions in W and Mo; elevated concentrations of refractory HSE) and more reducing conditions (i.e., elevated concentrations of Cr, Si, and C). Evidently, these signatures cannot be produced by a single event of metal-silicate fractionation.

The aforementioned evidence for formation of Nedagolla from differentiated material raises the question of whether some of the peculiar chemical characteristics of Nedagolla reflect the effects of fractional crystallization of its parental melt. For instance, Cr is incompatible in solid metal and is, therefore, expected to be enriched in the remaining liquid metal, ultimately resulting in higher Cr concentrations in late- compared to early-crystallized irons (Bonnand and Halliday 2018). However, in some magmatic groups Cr concentrations in early-crystallized irons are higher than in late-crystallized irons, suggesting that the Cr concentrations in iron meteorites are controlled by other processes, such as sequestration of chromite or sampling bias (Chabot et al. 2009). However, Nedagolla does not contain chromites, but Cr rather seems to be hosted in daubréelite, which is the host of Cr in metal formed under very reducing conditions (Wai and Wasson 1970). This combined with the evidence for reducing conditions inferred from the high Si content of Nedagolla indicates that the very high Cr concentration of Nedagolla compared to most other magmatic irons reflects neither of the aforementioned processes, but is a signature of metal segregation under stronger reducing conditions. Likewise, the W and Mo concentrations of different members of a given group of magmatic irons typically vary by less than a factor of ~2 (e.g., McCoy et al. 2011; Tornabene et al. 2020), indicating that core crystallization also cannot account for the depletions of W and Mo by a factor of ~2 and ~4 observed for Nedagolla. Thus, the W-Mo depletions and the Cr enrichment of Nedagolla must have been established prior to core crystallization and, therefore, reflect the bulk composition of the core material from which Nedagolla later derived.

Given the isotopic evidence that Nedagolla's formation involved mixing between NC and CC materials, it is tempting to attribute the chemical evidence for metal segregation under oxidizing and reducing conditions combined in a single meteorite to the same mixing process. Metal segregation in CC bodies is expected to occur under more oxidizing conditions, because these objects likely formed beyond or at the snow line and are, therefore, expected to have incorporated water ice. This is consistent with the core sizes inferred for iron meteorite parent bodies, which tend to be smaller for CC compared to NC bodies (Hilton et al. 2020; Rubin 2018; Tornabene et al. 2020). Conversely, metal segregation in NC bodies is expected to occur under more reducing conditions, because these objects formed in the inner solar system and, therefore, probably inside the snowline. Thus, mixing of two metals that previously segregated in NC and CC bodies may account for the chemical signatures observed for Nedagolla. We will return to this issue further below when discussing the isotopic evidence for mixing between NC and CC materials in more detail.

## Origin of mixed NC-CC isotopic composition

The Mo, Ru, and Ni isotopic composition of Nedagolla may be used to constrain the isotopic compositions of the NC and CC material that was mixed together to produce Nedagolla or its precursor. These data may also be used to assess the chemical composition of these materials by comparing the mass fractions of Mo, Ru, and Ni derived from NC and CC material, respectively. The mass fractions of CC-derived Mo, Ru, and Ni in Nedagolla can be calculated by mass balance as follows (Budde et al., 2019):

$$f_{CC} = \frac{(R_{Nedagolla} - R_{NC})}{(R_{CC} - R_{NC})} \quad (1)$$

where *R* is a Mo, Ru, or Ni isotopic ratio. If two bodies with chondritic relative proportions of Mo, Ru, and Ni were mixed, then the mass fraction of CC material in Nedagolla would be the same for all three elements. Evidently, this would not be the case if chemically fractionated material was involved in the mixing.

For Mo, the fraction of CC-derived material in Nedagolla can be calculated from its position between the NC- and CC-lines in the $\varepsilon^{95}$Mo–$\varepsilon^{94}$Mo diagram. This is because the NC- and CC-lines are approximately parallel, such that Nedagolla's composition divides any tie line between them into two segments whose ratio to each other is approximately constant (Budde et al. 2019). To this end, it is useful to employ the $\Delta^{95}$Mo notation, which is the parts-per-million deviation of a sample's Mo isotopic composition from a theoretical *s*-process mixing line passing through the origin (Budde et al. 2019):

$$\Delta^{95}\text{Mo} = (\varepsilon^{95}\text{Mo} - 0.596 \times \varepsilon^{94}\text{Mo}) \times 100. \quad (2)$$

The characteristic $\Delta^{95}$Mo value of the CC reservoirs is 26±2 and is derived from the *y*-axis intercept of the linear regressions of available CC meteorite data (Budde et al. 2019). It is important to recognize that the slope of the CC-line is in excellent agreement with the predicted slope of *s*-process variations. The Mo isotope variations along the CC-line can, therefore, be accounted for by pure *s*-process variations, meaning that the $\Delta^{95}$Mo value of 26±2 is a characteristic value for all CC meteorites (Budde et al. 2019; Kleine et al. 2020). For the NC reservoir the situation is more complicated, because recent work has shown that the slope of the NC-line is slightly shallower than the expected slope of a pure *s*-process mixing line (Spitzer et al. 2020b). As such, there is not a single characteristic $\Delta^{95}$Mo value for all NC meteorites, which instead depends on the position of a given sample along the NC-line. However, the effect of these non-uniform $\Delta^{95}$Mo values among NC meteorites on the overall uncertainty of the calculated CC mass fractions is comparatively small. For instance, assuming the nominal $\Delta^{95}$Mo = -9±2 for the NC endmember (Budde et al., 2019), and $\Delta^{95}$Mo = 3±5 for Nedagolla determined in this study, results in a CC-derived Mo fraction in Nedagolla of 33±16 %. This value changes to 30±18 % if the typical $\Delta^{95}$Mo = -7±3 (Render et al. 2017) for enstatite chondrites is used instead.

For Ru and Ni, it is more difficult to quantify the relative contributions of NC and CC material, because the inferred fraction of CC material will always depend on the assumed composition of the NC and CC mixing endmembers. This is illustrated in Fig. 6, which shows the inferred CC fractions for Ru and Ni as a function of the assumed isotopic composition of the CC endmember. As is evident from the $\varepsilon^{95}$Mo–$\varepsilon^{94}$Mo diagram, any tie line between the NC- and CC-lines that passes through Nedagolla's composition and connects known bulk NC and CC meteorites, intersects the NC-line close to the composition of enstatite chondrites (Fig. 3). This suggests strongly that the NC mixing endmember had an enstatite chondrite-like isotopic composition and is consistent with the aforementioned chemical evidence for formation under reducing conditions, such as the high Si concentration of Nedagolla metal, which is also observed for enstatite chondrites. In the mixing calculations for Ru and Ni we, therefore, assumed an enstatite chondrite-like isotopic composition for the NC endmember.

With this assumption, the Ni isotopic composition of Nedagolla corresponds to a mass fraction of CC-derived Ni of between ~0.2 and ~0.6, which is consistent with the mass fraction of CC-derived Mo determined above. By contrast, for a typical Ru isotopic composition of CC iron meteorites ($\varepsilon^{100}$Ru ≈ –1; Worsham et al., 2019), the inferred fraction of CC-derived Ru in Nedagolla is only ~10% (Fig. 6). However, as noted above, the CC endmember may be responsible for the elemental Mo depletion relative to Ru in Nedagolla, which is a chemical signature of metal segregation under more oxidizing conditions in CC bodies. In this case, the fraction of CC-derived Mo in Nedagolla should be lower than the fraction of CC-derived Ru, because a CC body with subchondritic Mo/Ru contributed less Mo than Ru to the final mixture. As the mass fraction of CC-derived Mo in Nedagolla is fixed at 30±18 % (as inferred using an enstatite chondrite-like $\Delta^{95}$Mo, see above), this would imply that the mass fraction of CC-derived Ru in Nedagolla should be higher than this value. The mass balance calculations shown in Fig. 6 would then imply that the $\varepsilon^{100}$Ru value of the CC endmember was larger than about –0.8.

The inference of $\varepsilon^{100}$Ru > –0.8 for the CC material in Nedagolla is a somewhat surprising result, because most CC meteorites have more negative $\varepsilon^{100}$Ru values. This is especially true for CC irons, all of which are characterized by an $\varepsilon^{100}$Ru value of about –1 (e.g., Worsham et al. 2019). There are a few carbonaceous chondrites, such as CM and CI chondrites, with more variable and smaller anomalies, but this has been attributed, at least in part, to modifications of Ru isotope signatures by parent body processes, which created variations among different members of a given group of carbonaceous chondrites (Fischer-Gödde and Kleine 2017). As such, the characteristic $\varepsilon^{100}$Ru of bulk carbonaceous chondrite parent bodies, and whether they are different from the value of about –1 observed for CC irons, are unknown.

Further insights into the composition of the CC endmember that contributed to Nedagolla may be obtained by also considering anomalies in $\varepsilon^{60}$Ni. Whereas the $\varepsilon^{62}$Ni and $\varepsilon^{64}$Ni anomalies of Nedagolla overlap with those of CC meteorites, its $\varepsilon^{60}$Ni is distinct and overlaps with the typical value of NC meteorites. This is illustrated in the $\varepsilon^{62}$Ni–$\varepsilon^{60}$Ni diagram, where Nedagolla plots on the extension of the NC trend, above the CC field (Fig. 4b). As such, any mixture between enstatite chondrites and typical CC meteorites cannot reproduce the $\varepsilon^{62}$Ni–$\varepsilon^{60}$Ni systematics of Nedagolla. Variations in $\varepsilon^{60}$Ni may in part also have a radiogenic origin, reflecting early Fe-Ni fractionation and subsequent decay of $^{60}$Fe. Specifically, the silicate mantles of differentiated objects are characterized by high Fe/Ni ratios, and so if core formation occurred during the lifetime of $^{60}$Fe, then the mantle may evolve to radiogenic $^{60}$Ni anomalies. However, owing to the rather low solar system initial $^{60}$Fe/$^{56}$Fe ratio of ~$1\times10^{-8}$ (Tang and Dauphas 2012), radiogenic $^{60}$Ni variations are generally small. For instance, the bulk silicate portion of the eucrite parent body has an estimated $\varepsilon^{60}$Ni excess of only ~0.23, and so, given its rather low Ni content (~100 µg/g; Warren et al. 1999) compared to the high Ni content of the core (~8 wt.% Ni), addition of mantle-derived, radiogenic $^{60}$Ni to the core has no measurable effect on the final $\varepsilon^{60}$Ni of the core. As such, Nedagolla's more elevated $\varepsilon^{60}$Ni compared to most CC meteorites does not result from metal-silicate re-equilibration during the NC-CC mixing.

CI chondrites are the only CC meteorites for which more elevated $\varepsilon^{60}$Ni have been reported (Regelous et al. 2008), and it is noteworthy that the Ni isotopic composition of Nedagolla is consistent with a mixture between materials with enstatite and CI chondrite-like isotopic compositions (Fig. 4). As noted above, this mixture could also account for the observed Ru isotopic composition of Nedagolla, such that, based on available data, Nedagolla's isotopic composition is best reproduced as a mixture between two bodies with enstatite chondrite- and CI-like isotopic compositions (Fig. 6, 7). In this case, both bodies were mixed in about equal proportions (i.e., $f_{CC} \approx 0.5$ for Ru) and the CC contribution for Mo was smaller, because the CC body was depleted in Mo relative to Ru. However, although this mixing model seems to be consistent with several chemical and isotopic characteristics of Nedagolla, it heavily relies on the assumption that the measured Ru and Ni isotopic compositions of CI chondrites are representative of their bulk parent body, which is unclear at present. As such, an improved understanding of the characteristic bulk isotopic compositions of carbonaceous chondrite parent bodies will be necessary for better constraining the nature of the CC material that contributed to Nedagolla.

Regardless of its isotopic composition, the CC body likely was a metallic object, as is evident from the Mo and W depletions relative to other refractory siderophile elements in Nedagolla, which most likely reflect metal segregation under more oxidizing conditions. However, metal segregation under such conditions after mixing of the NC and CC bodies cannot produce the elevated Si and Cr contents of the Nedagolla metal, which require more reducing conditions. Conversely, addition of CC metal with the characteristic Mo and W depletions to a reduced NC body followed by metal segregation under reducing conditions can account for the observed chemical signatures, because the Mo-W depletions would be inherited from a previous metal segregation event. This requires, however, that the silicate portion of the CC body did not significantly participate to the mixing, because otherwise the Mo-W depletions would be erased through re-equilibration between the CC silicates and the CC metal under reducing conditions. It is also possible that two metallic bodies were mixed, where metal segregation in the CC body had occurred under more oxidizing and in the NC body under more reducing conditions. This would naturally result in a combined metallic object with chemical signatures indicating both oxidizing and reducing conditions in a single sample.

In summary, the mixed NC-CC isotopic composition of Nedagolla can be reproduced by mixing between materials with enstatite chondrite-like and CI chondrite-like isotopic compositions, with the caveat that it is unknown at present as to whether the isotopic composition measured for CI chondrites is representative for bulk CC parent bodies. At least the CC object likely was a differentiated body, and only the metallic core of this body contributed to the mixing that ultimately produced Nedagolla. By contrast, the NC body may have been a differentiated or an undifferentiated object.

## Hf-W chronology

The $^{182}W$ composition of Nedagolla provides two important constraints about the mixing processes that produced its parental melt. First, the more radiogenic $\varepsilon^{182}W$ value of Nedagolla compared to other magmatic iron meteorites suggests that the formation of Nedagolla involved some later re-equilibration with silicate material that is characterized by elevated Hf/W and, hence, radiogenic $^{182}W$. This re-equilibration may have occurred during the NC-CC mixing event or, alternatively, during an earlier impact event on one of the bodies involved in this event. However, given that all other magmatic irons investigated so far have less radiogenic $^{182}W$ compositions than Nedagolla, such metal-silicate re-equilibration events on iron meteorite parent bodies appear to be rare, and so it seems more likely that the more radiogenic $^{182}W$ of Nedagolla reflects metal-silicate re-equilibration during the NC-CC mixing event. Second, the $\varepsilon^{182}W$ value of Nedagolla can be used to assess the timing of NC-CC mixing and metal segregation on its parent body. Specifically, a model age of Hf-W fractionation from an unfractionated chondritic reservoir can be calculated as follows (e.g., Horan et al. 1998):

$$\Delta t_{CAI} = -\frac{1}{\lambda}\ln\left[\frac{(\varepsilon^{182}W)_{sample} - (\varepsilon^{182}W)_{chondrites}}{(\varepsilon^{182}W)_{SSI} - (\varepsilon^{182}W)_{chondrites}}\right] \quad (3)$$

where $(\varepsilon^{182}W)_{chondrites}$ is the composition of carbonaceous chondrites (-1.91 ± 0.08) (Kleine et al. 2009), $(\varepsilon^{182}W)_{SSI}$ is the Solar System initial (-3.49 ± 0.07) obtained from CAI (Kruijer et al. 2014a), and λ is the $^{182}Hf$ decay constant of 0.0778 ± 0.0015 $Ma^{-1}$ (Vockenhuber et al. 2004). The resulting Hf-W model age for Nedagolla is 7±2 Ma after CAI formation.

The Hf-W model age is chronologically meaningful only if the metallic melt from which Nedagolla crystallized equilibrated within a reservoir characterized by a bulk chondritic Hf-W ratio and $^{182}W$ composition. Given that Nedagolla's formation involved mixing between an NC and a CC body, this would require that the silicate portions of both bodies fully equilibrated with their respective metal. However, as noted above, it appears that at least the CC body was chemically differentiated and its silicate portion was most likely not involved in the mixing process, because otherwise the characteristic Mo-W depletions observed for Nedagolla would likely not be preserved. As such, the final parental metallic melt of Nedagolla can have equilibrated only with the silicate portion of the NC body. Thus, even if this equilibration was complete, the parental melt of Nedagolla nevertheless segregated from a reservoir with subchondritic silicate-to-metal ratio. In this case, and also if the equilibration with the silicate portion of the NC body was incomplete, will the Hf-W model age always predate the true time of mixing (Kruijer and Kleine 2019). Thus, the collision that produced the parental melt of Nedagolla likely occurred later than 7±2 Ma after CAI formation, but its exact timing remains unconstrained.

## Formation model

The discussion up to this point has shown that the parental melt of Nedagolla most likely formed by mixing between CC metal and NC metal, which at least partially re-equilibrated with NC silicates. This mixing occurred relatively late and must, therefore, be related to the collision of an NC body and a CC body. As noted above, the isotopic systematics do not allow determining whether the NC body was differentiated or not. However, the bulk chemical composition of Nedagolla suggests that it more likely formed by collision between two differentiated objects. This is because it is difficult to envision how Nedagolla's extreme depletion in volatile elements should have been produced by impact onto an undifferentiated NC body. Formation of Nedagolla's parental melt in this manner would require the subsequent segregation of the metallic melt and efficient degassing from this melt. The formation of IAB and IIE irons probably involved impact-related processes that induced melting as well metal-silicate re-equilibration and mixing (Hilton and Walker 2020; Kruijer and Kleine 2019; Wasson 2017; Wasson and Kallemeyn 2002; Wasson and Wang 1986). However, these irons are not particularly volatile element-depleted, and so it seems unlikely that Nedagolla lost its volatiles by such a process. Moreover, the IAB and IIE irons contain abundant silicate inclusions, but Nedagolla does not. Together, these observations suggest that Nedagolla did not form by impact of a CC core onto an undifferentiated NC body.

A more realistic pathway to account for Nedagolla's extreme volatile depletion and the mixing of NC and CC materials by the same process is to invoke a hit-and-run collision between two differentiated bodies (Asphaug et al. 2006). Such a collision may result in removal of the silicate mantle from the smaller of the two colliding bodies, and it may also result in mixing of the released core material of the smaller body with mantle and core material of the larger body. Moreover, removal of the overlying silicate mantle results in depressurization of the still molten metallic core, which may lead to degassing. As such, this scenario can likely account for many of the isotopic and chemical characteristics of Nedagolla. Our preferred model thus is that the parent body of Nedagolla formed as the result of a hit-and-run collision between two differentiated bodies of different sizes, where the smaller object was a CC body and the larger object an NC body. As is evident from its extremely rapid cooling rate, Nedagolla itself derives from the near surface area of its parent body, where it may have been re-melted during a later, small impact event (Miyake and Goldstein 1974). This event, however, likely had no significant effect on the isotopic or chemical composition of Nedagolla.

## Conclusions

The ungrouped iron meteorite Nedagolla is the first meteorite with a bulk isotopic composition intermediate between those of the non-carbonaceous and carbonaceous meteorite reservoirs. The mixed NC-CC isotopic signature of Nedagolla is particularly evident from its Mo isotopic composition, but also from the combined Mo-Ru-Ni isotope systematics, where Nedagolla does not plot consistently in either the NC or CC compositional fields. The $^{182}W$ isotopic composition of Nedagolla indicates that its mixed NC-CC isotopic composition has been established relatively late, more than ~7 Ma after Solar System formation. As such, Nedagolla most likely is the result of a collision between a NC and a CC body. Apart from its mixed NC-CC isotopic signature, Nedagolla's chemical composition suggests mixing of metal formed under reducing conditions as likely prevailed in some NC bodies with metal formed under the more oxidizing conditions characteristic for core formation in CC bodies. Together, these observations suggest that the parent body of Nedagolla formed as the result of a collision between two differentiated, isotopically and chemically distinct objects.

Although NC and CC bodies formed in two distinct areas of the disk, which most likely were separated by Jupiter, they occur together in the present-day asteroid belt. This is thought to be a natural outcome of the growth and/or migration of the gas giant planets, which result in inward scattering of CC bodies into the inner Solar System (Raymond and Izidoro 2017). This process is expected to result in energetic collisions among NC and CC bodies, but direct empirical evidence for such collisions has thus far been lacking. To this end, Nedagolla may be viewed as the first documented evidence for the energetic collisions between NC and CC bodies that are expected to accompany the implantation of CC bodies into the asteroid belt.

# Acknowledgements

We are deeply honored to contribute to this special issue dedicated to the memory of John Wasson, who has made so many fundamental contributions to our understanding of iron meteorites and their importance for constraining the earliest history of the solar system. We are grateful to Natasha Almeida (National History Museum, London) for providing a sample of Nedagolla for this study. Constructive reviews by Alan Rubin, Richard Walker, and associate editor Ed Scott are gratefully acknowledged. Funded by the Deutsche Forschungsgemeinschaft (DFG, German Research Foundation) – Project-ID 263649064 – TRR 170. This is TRR 170 pub. no. 139.

Table 1. Pt isotope data for Nedagolla and NIST 129c.

| sample | ID | N | $\varepsilon^{192}$Pt (6/5)[a] | $\varepsilon^{194}$Pt (6/5)[a] | $\varepsilon^{198}$Pt (6/5)[a] | $\varepsilon^{192}$Pt (8/5)[a] | $\varepsilon^{194}$Pt (8/5)[a] | $\varepsilon^{196}$Pt (8/5)[a] |
|---|---|---|---|---|---|---|---|---|
| Nedagolla | UI-05 | 4 | 0.8 ± 1.1 | 0.50 ± 0.13 | -0.55 ± 0.15 | 0.1 ± 1.3 | 0.33 ± 0.05 | 0.18 ± 0.05 |
| Nedagolla | UI-05[b] | 3 | 1.3 ± 1.2 | 0.33 ± 0.18 | -0.30 ± 0.20 | 1.0 ± 1.1 | 0.25 ± 0.15 | 0.10 ± 0.07 |
| weighted mean | | | 1.1 ± 0.8 | 0.45 ± 0.11 | -0.46 ± 0.12 | 0.61 ± 0.85 | 0.32 ± 0.05 | 0.16 ± 0.04 |
| NIST 129c | | 32 | 0.05 ± 0.21 | 0.03 ± 0.03 | -0.04 ± 0.04 | 0.02 ± 0.21 | 0.02 ± 0.03 | 0.01 ± 0.01 |

[a] Normalized to $^{196}$Pt/$^{195}$Pt = 0.7464 ('6/5') and $^{198}$Pt/$^{195}$Pt = 0.2145 ('8/5') and reported uncertainties are 95% confidence intervals.

Table 2. Mo isotope data for Nedagolla and NIST 129c.

| sample | ID | N | $\varepsilon^{92}Mo_{meas}$[a] | $\varepsilon^{94}Mo_{meas}$[a] | $\varepsilon^{95}Mo_{meas}$[a] | $\varepsilon^{97}Mo_{meas}$[a] | $\varepsilon^{100}Mo_{meas}$[a] | $\varepsilon^{92}Mo_{CRE-corr}$[a] | $\varepsilon^{94}Mo_{CRE-corr}$[a] | $\varepsilon^{95}Mo_{CRE-corr}$[a] | $\varepsilon^{97}Mo_{CRE-corr}$[a] | $\varepsilon^{100}Mo_{CRE-corr}$[a] | $\Delta^{95}$Mo[b] |
|---|---|---|---|---|---|---|---|---|---|---|---|---|---|
| Nedagolla | UI-05 | 8 | 0.66 ± 0.05 | 0.54 ± 0.07 | 0.32 ± 0.06 | 0.15 ± 0.04 | 0.14 ± 0.07 | 0.74 ± 0.06 | 0.58 ± 0.07 | 0.37 ± 0.06 | 0.16 ± 0.05 | 0.12 ± 0.07 | 2.7 ± 7.7 |
| Nedagolla | UI-05b | 8 | 0.61 ± 0.16 | 0.55 ± 0.08 | 0.32 ± 0.06 | 0.17 ± 0.08 | 0.07 ± 0.05 | 0.68 ± 0.16 | 0.60 ± 0.08 | 0.38 ± 0.06 | 0.19 ± 0.08 | 0.05 ± 0.05 | 2.6 ± 7.9 |
| weighted mean | | | | | | | | 0.73 ± 0.05 | 0.59 ± 0.06 | 0.38 ± 0.04 | 0.17 ± 0.04 | 0.08 ± 0.04 | 2.7 ± 5.4 |
| NIST 129c[c] | | 38 | -0.27 ± 0.05 | -0.09 ± 0.03 | 0.00 ± 0.02 | 0.03 ± 0.02 | -0.17 ± 0.03 | | | | | | |

[a] Normalized to $^{98}$Mo/$^{96}$Mo = 1.453173 and reported uncertainties are 95% confidence intervals.

[b] Calculated using the equation provided in Budde et al. (2019): $\Delta^{95}Mo = (\varepsilon^{95}Mo - 0.596 \times \varepsilon^{94}Mo) \times 100$.

[c] The Mo isotopic compostion of the industrially produced steel standard NIST SRM 129c has been shown to be affected by non-expinential mass fractionation, resulting in anomalous (lower) $\varepsilon^{i}$Mo values (Burkhardt et al. 2011, Budde et al. 2019).

Table 3. Ru isotope data for Nedagolla and doped reference materials.

| sample | ID | N | $\varepsilon^{96}Ru_{meas}$[a] | $\varepsilon^{98}Ru_{meas}$[a] | $\varepsilon^{100}Ru_{meas}$[a] | $\varepsilon^{102}Ru_{meas}$[a] | $\varepsilon^{104}Ru_{meas}$[a] | $\varepsilon^{96}Ru_{CRE-corr}$[a] | $\varepsilon^{98}Ru_{CRE-corr}$[a] | $\varepsilon^{100}Ru_{CRE-corr}$[a] | $\varepsilon^{102}Ru_{CRE-corr}$[a] | $\varepsilon^{104}Ru_{CRE-corr}$[a] |
|---|---|---|---|---|---|---|---|---|---|---|---|---|
| Nedagolla | UI-05c | 6 | 0.33 ± 0.14 | 0.00 ± 0.48 | -0.09 ± 0.07 | 0.03 ± 0.05 | 0.23 ± 0.08 | 0.22 ± 0.15 | -0.07 ± 0.48 | -0.17 ± 0.08 | -0.03 ± 0.08 | 0.11 ± 0.10 |
| standards[b] | | 84 | 0.06 ± 0.05 | 0.02 ± 0.06 | -0.01 ± 0.01 | 0.02 ± 0.02 | 0.08 ± 0.03 | | | | | |

[a] Normalized to $^{99}$Ru/$^{101}$Ru = 0.7450754 and reported uncertainties are 95% confidence intervals.

[b] Based on previous studies from this laboratory (Fischer-Gödde et al. 2015).

Table 4. Ni isotope data for Nedagolla and NIST 361.

| sample | ID | N | $\varepsilon^{60}$Ni (61/58)[a] | $\varepsilon^{62}$Ni (61/58)[a] | $\varepsilon^{64}$Ni (61/58)[a] | $\varepsilon^{58}$Ni (62/61)[a] | $\varepsilon^{60}$Ni (62/61)[a] | $\varepsilon^{64}$Ni (62/61)[a] |
|---|---|---|---|---|---|---|---|---|
| Nedagolla | UI-05 | 20 | 0.00 ± 0.03 | 0.09 ± 0.04 | 0.22 ± 0.09 | 0.28 ± 0.14 | 0.09 ± 0.07 | -0.05 ± 0.11 |
| Nedagolla | UI-05c | 18 | 0.01 ± 0.03 | 0.11 ± 0.09 | 0.16 ± 0.12 | 0.34 ± 0.27 | 0.12 ± 0.12 | -0.17 ± 0.16 |
| weighted mean | | | 0.01 ± 0.02 | 0.10 ± 0.04 | 0.20 ± 0.07 | 0.29 ± 0.12 | 0.10 ± 0.06 | -0.09 ± 0.09 |
| NIST 361 | | 64 | -0.01 ± 0.02 | 0.03 ± 0.03 | 0.10 ± 0.07 | 0.08 ± 0.10 | 0.02 ± 0.05 | 0.02 ± 0.06 |

[a] Normalized to $^{61}$Ni/$^{58}$Ni = 0.016744 ('61/58') and $^{62}$Ni/$^{61}$Ni = 3.1884 ('62/61') and reported uncertainties are 95% confidence intervals.

Table 5. W isotope data for Nedagolla, NIST129c, and NIST 361.

| sample | ID | N | $\varepsilon^{182}W (6/4)_{meas}$[a] | $\varepsilon^{183}W (6/4)_{meas}$[a] | $\varepsilon^{182}W (6/4)_{nuc-corr}$[a] | $\varepsilon^{182}W (6/4)_{CRE-corr}$[a] | $\Delta t_{CAI}$ (Ma) | $\varepsilon^{182}W (6/3)_{meas}$[a] | $\varepsilon^{184}W (6/3)_{meas}$[a] | $\varepsilon^{182}W (6/3)_{nuc-corr}$[a] | $\varepsilon^{182}W (6/3)_{CRE-corr}$[a] | $\Delta t_{CAI}$ (Ma) |
|---|---|---|---|---|---|---|---|---|---|---|---|---|
| Nedagolla[b] | UI-05 | 5 | -2.85 ± 0.11 | 0.09 ± 0.09 | -2.97 ± 0.17 | -2.77 ± 0.18 | 7.8 ± 2.8 | -3.00 ± 0.11 | -0.06 ± 0.06 | -3.01 ± 0.11 | -2.81 ± 0.12 | 7.3 ± 1.9 |
| NIST 129c, NIST 361[b] | | 17 | 0.03 ± 0.02 | 0.03 ± 0.02 | | | | 0.04 ± 0.02 | -0.02 ± 0.01 | | | |

[a] Normalized to $^{186}$W/$^{184}$W = 0.92767 ('6/4') and $^{186}$W/$^{183}$W = 1.98590 ('6/3') and reported uncertainties are 95% confidence intervals.

[b] Reported measured isotope ratios involving $^{183}$W have already been corrected for the mass-independent effect (see methods).

Table 6. Chemical data for Nedagolla.

| Element | 50% $T_C$ [K] | Literature[a] | This study[b] |
| --- | --- | --- | --- |
| Os [ppb] | 1806 | 5827 | n.d. |
| Re [ppb] | 1736 | 473 | 424 |
| W [ppb] | 1736 | | 594 |
| Ir [ppb] | 1566 | 5165 | 4474 |
| Ru [ppb] | 1533 | 6266 | 5798 |
| Mo [ppb] | 1520 | | 2455 |
| Pt [ppb] | 1370 | 8840 | 7952 |
| Rh [ppb] | 1370 | | 944 |
| Ni [wt%] | 1363 | 6.0 | 6.1 |
| Co [wt%] | 1354 | 0.31 | 0.36 |
| Fe [wt%] | 1338 | 93.4 | 93.5 |
| Pd [ppb] | 1330 | 1883 | 1753 |
| Si [wt%] | 1314 | 0.14 | n.d. |
| Cr [wt%] | 1291 | 0.26 | 0.26 |
| P [wt%] | 1287 | 0.02 | n.d. |
| Cu [ppm] | 1034 | 1.5 | n.d. |
| Ga [ppb] | 1010 | 650 | 765 |
| Au [ppb] | 967 | 220 | n.d. |
| Ge [ppb] | 830 | 5 | n.d. |
| Zn [ppm] | 704 | <1 | 3 |
| C [wt%] | 40 | ~ 0.1 | n.d. |

[a] Literature data from Walker et al. (2005), Crocket (1972), Wai and Wasson (1970), and the Meteoritical Bulletin.

[b] Measured using Q-ICPMS relative to a multi-element standard solution. Estimated uncertainty is ~10% (RSD).

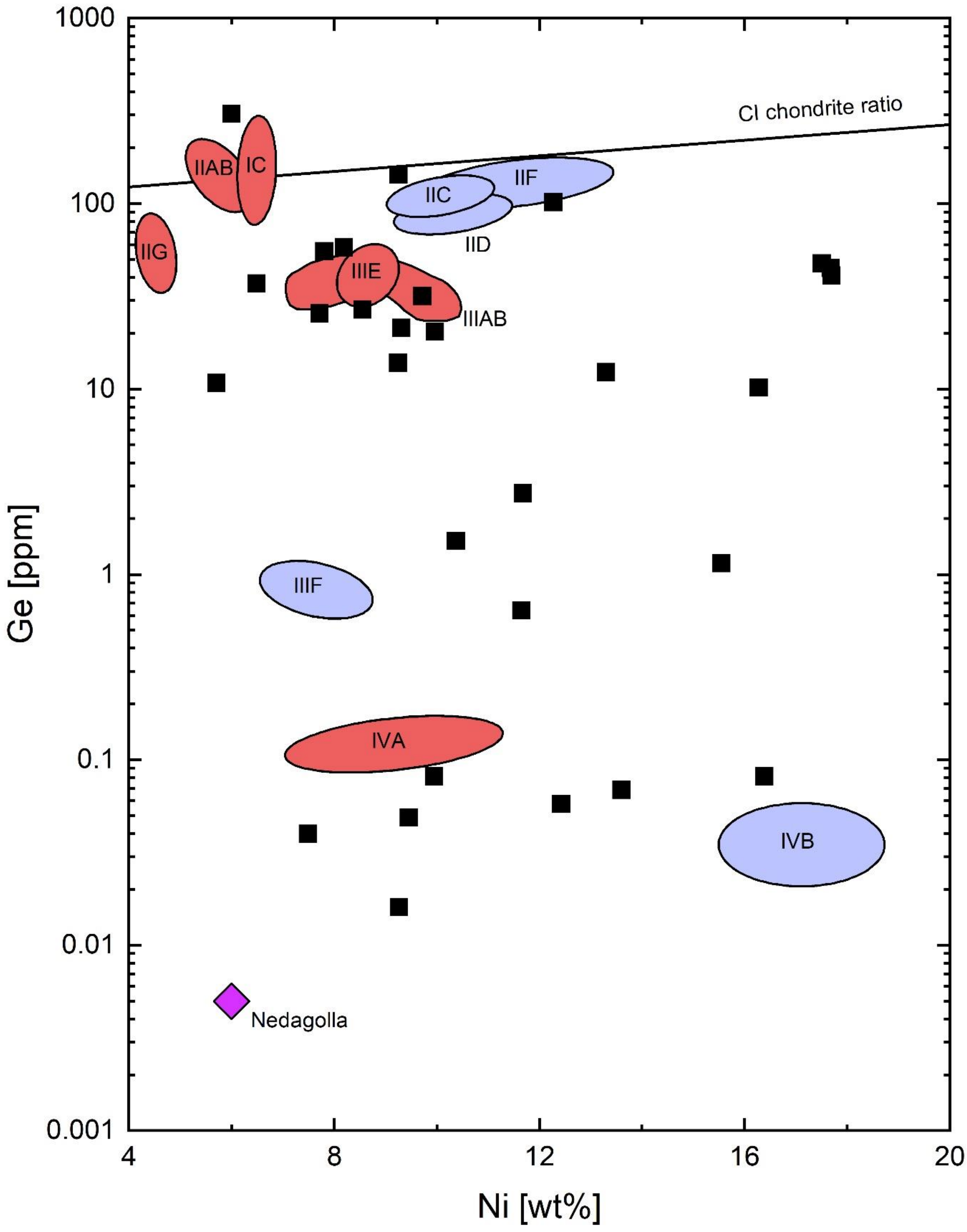

Fig. 1. Diagram of Ge versus Ni concentration for the major magmatic iron meteorite groups (Scott and Wasson, 1975). Red color indicates NC and blue color CC origin. Also shown are Nedagolla (purple diamond) and some selected ungrouped irons (black squares) for comparison. Note that Nedagolla has the lowest Ge concentration known among iron meteorites, which is evidence for an extreme depletion in volatile elements.

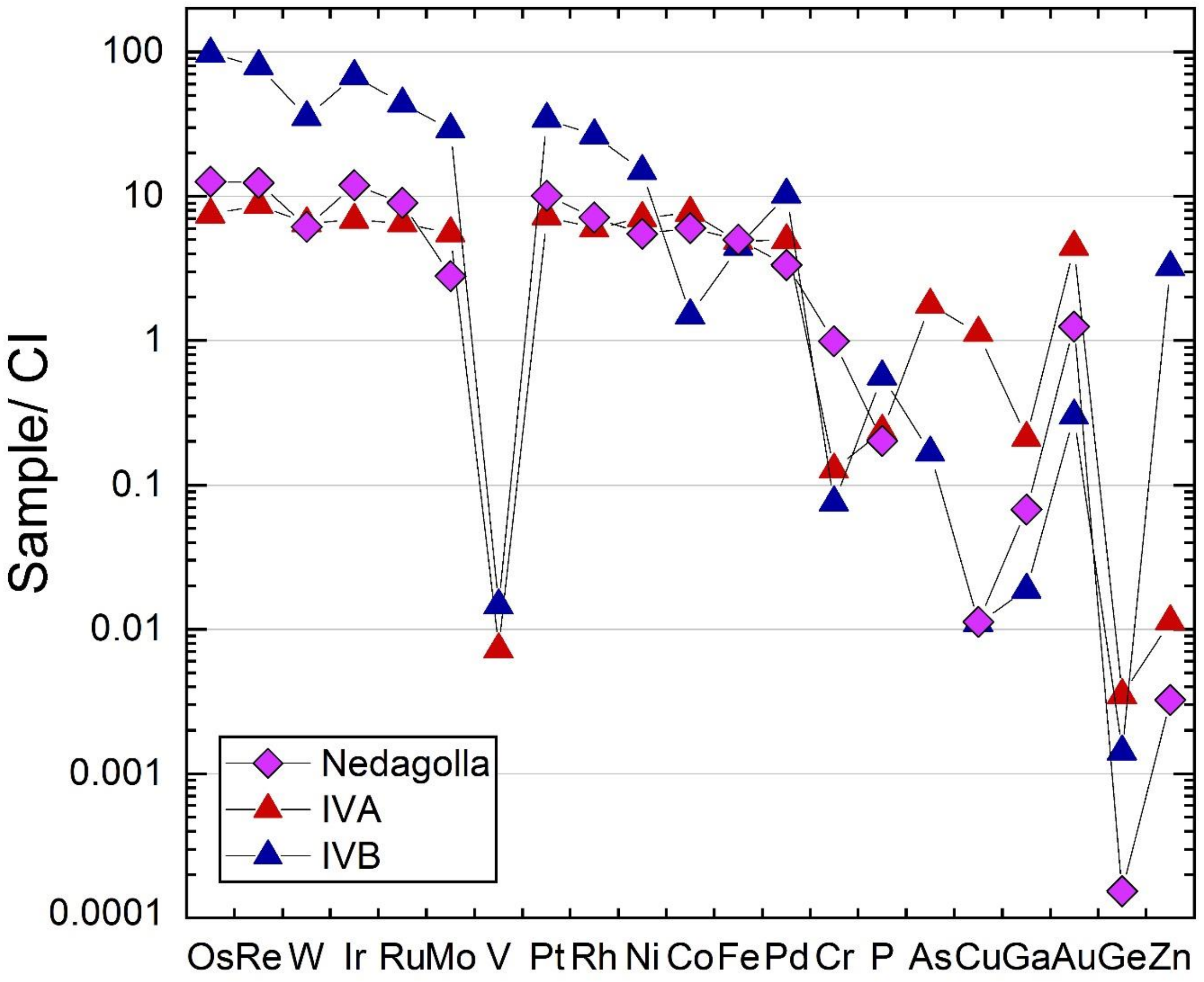

Fig. 2. CI chondrite-normalized abundances of siderophile elements for Nedagolla. The elements are plotted in order of decreasing 50% condensation temperature (Wood et al. 2019). Also shown for comparison is the average composition of early crystallized IVA (Jamestown, Marie Elena, La grange, Gibeon; McCoy et al., 2011) and IVB (Cape of Good Hope, Hoba, Tlacotepec; Walker et al., 2008) irons, which are the most volatile-depleted NC and CC iron meteorite groups. Data for Nedagolla as provided in Table 6; CI chondrite data as compiled by Yoshizaki and McDonough (2020).

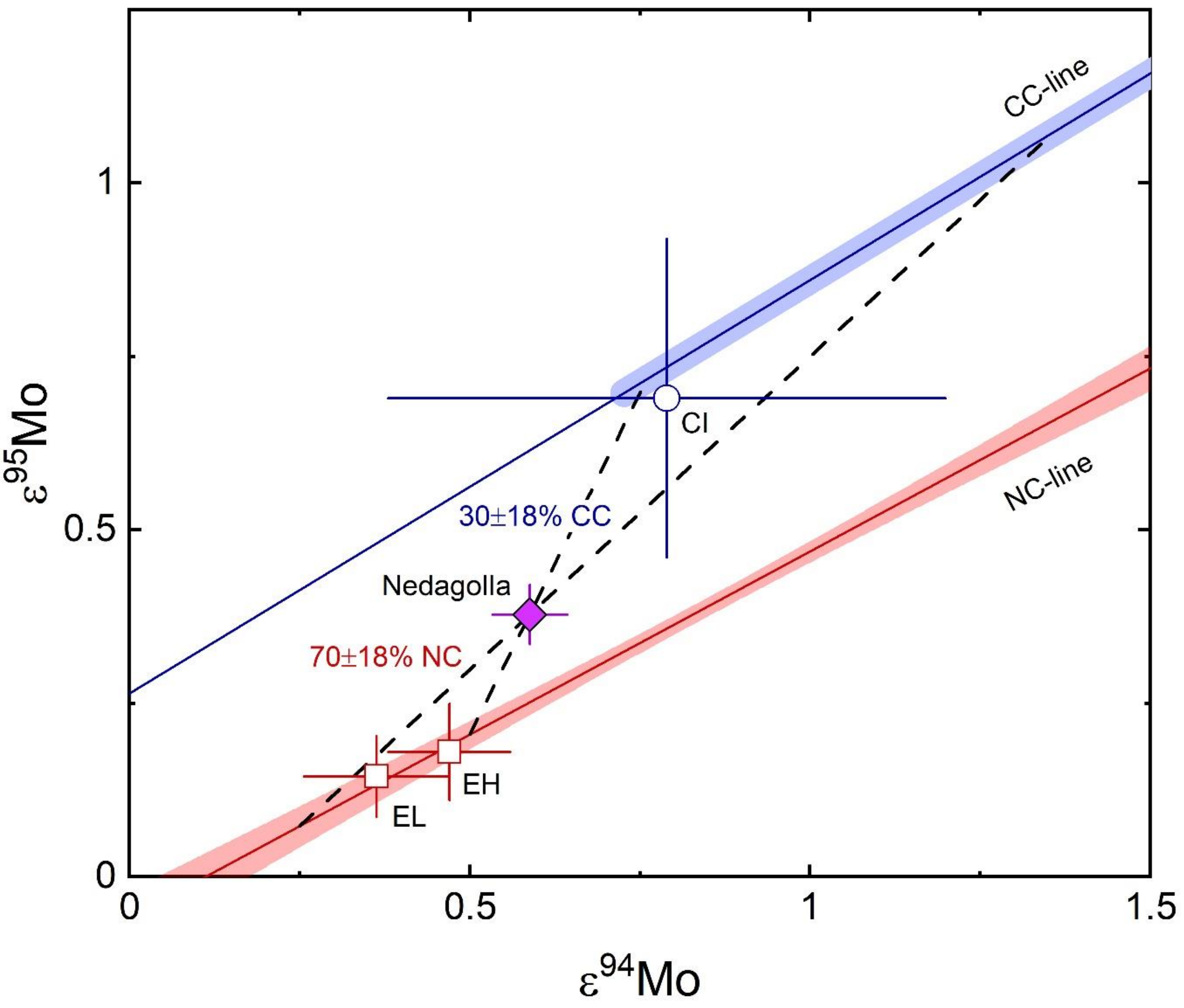

Fig. 3. Diagram of $\varepsilon^{95}$Mo versus $\varepsilon^{94}$Mo for Nedagolla and selected meteorite groups. NC- and CC-lines are from linear regressions of available bulk meteorite data as described in Spitzer et al. (2020) and Budde et al. (2019), respectively. The shaded area represents the range of bulk meteorite compositions and the width corresponds to the uncertainties around the NC- and CC-lines. Mixing lines (black dashed lines) connecting the compositions of most known CC meteorites and Nedagolla intersects the NC line around the composition of enstatite chondrites.

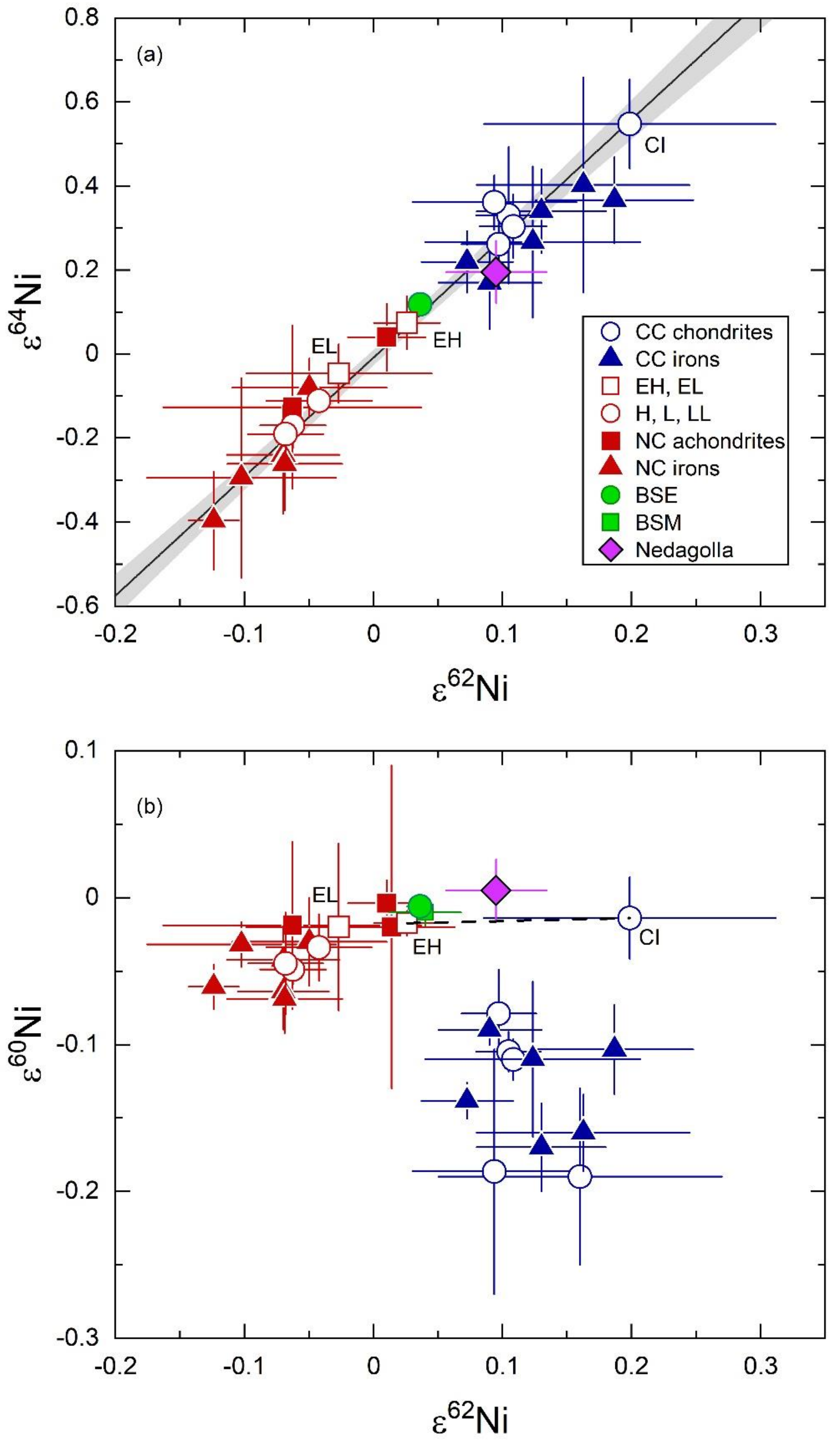

Fig. 4. Plots of (a) $\varepsilon^{64}$Ni vs. $\varepsilon^{62}$Ni, and (b) $\varepsilon^{60}$Ni vs. $\varepsilon^{62}$Ni. Data for NC and CC meteorites are from Burkhardt et al. (2017), Cook et al. (2020), Nanne et al. (2019), Regelous et al. (2008), Steele et al. (2011, 2012), and Tang and Dauphas (2012, 2014). In (a) Nedagolla plots at the lower end of the CC field, whereas in (b) it rather seems to plot along the extension of the trend defined by NC meteorites towards more positive $\varepsilon^{60}$Ni and $\varepsilon^{62}$Ni values. The dashed line in (b) represents a mixing line between EH and CI chondrites. BSM – bulk silicate Mars.

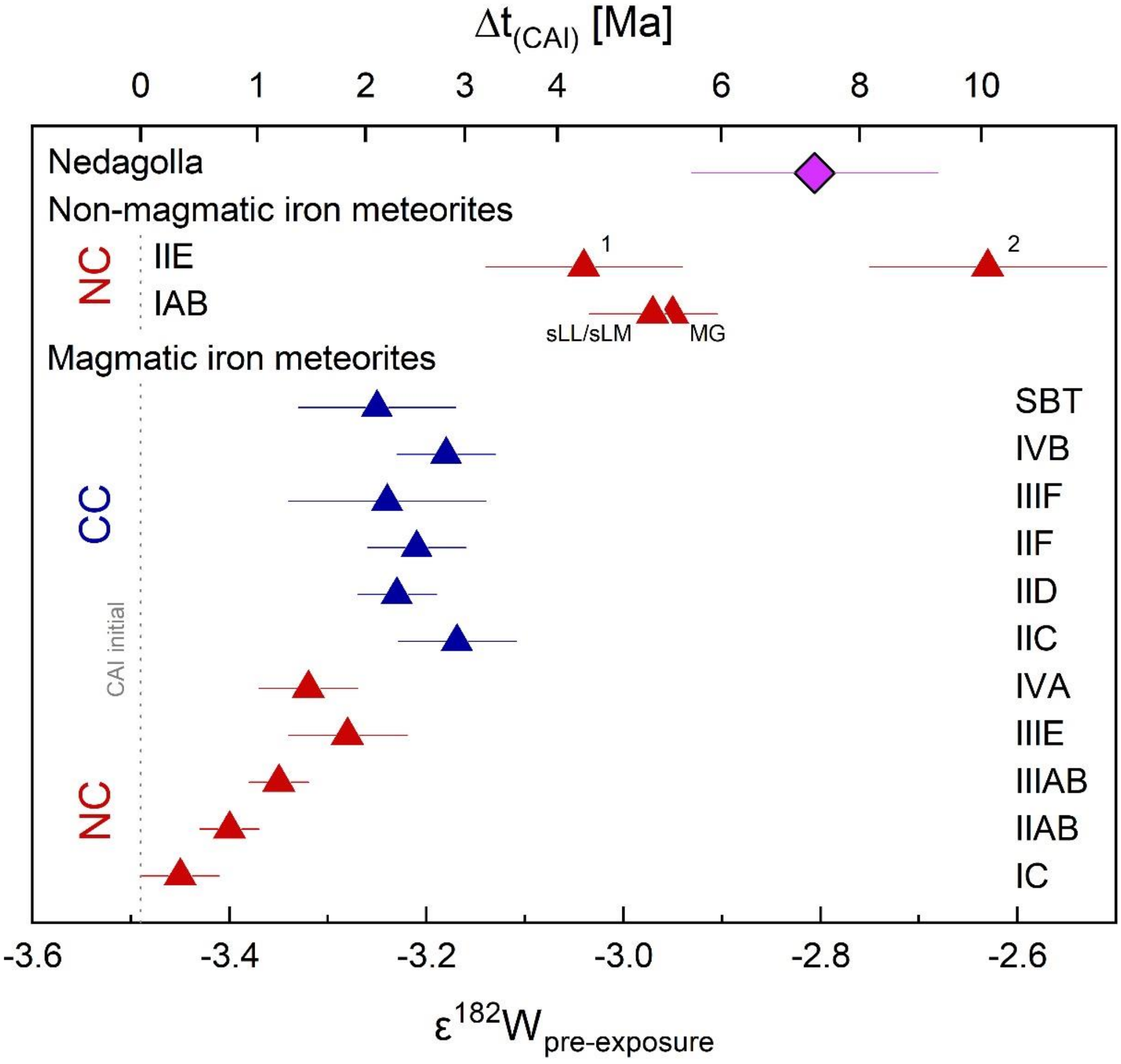

Fig. 5. Pre-exposure $\varepsilon^{182}W$ values (i.e., corrected for CRE exposure) and corresponding Hf-W model ages for various groups of iron meteorites. Data from Hilton and Walker (2020), Hilton et al. (2019), Hunt et al. (2018), Kruijer and Kleine (2019), Kruijer et al. (2014b, 2017), and Worsham et al. (2017). The $\varepsilon^{182}W$ value of Nedagolla is more radiogenic than values of other magmatic irons and overlaps with the composition of non-magmatic irons. The more elevated $\varepsilon^{182}W$ indicates later metal-silicate re-equilibration, most likely during impact processes.

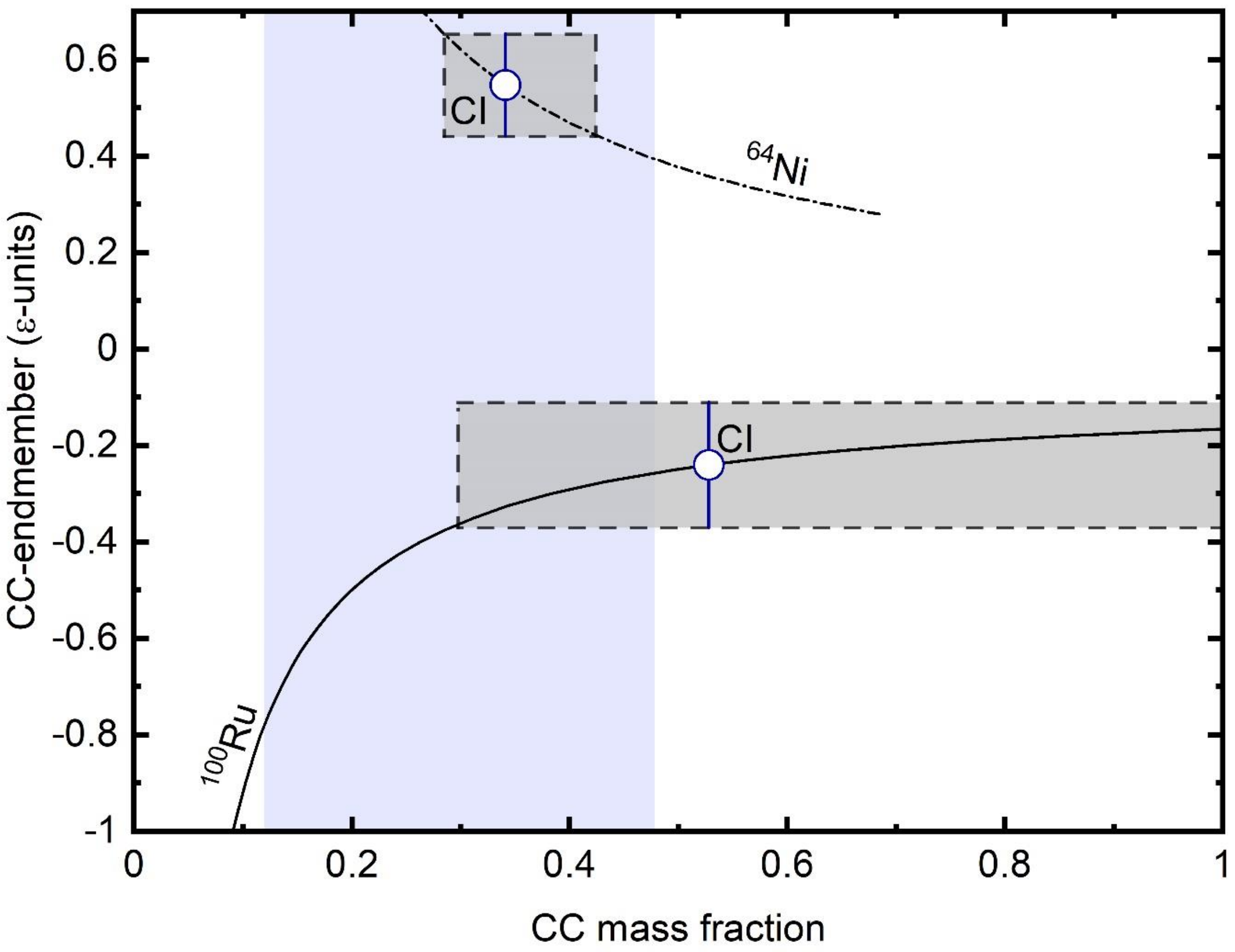

Fig. 6. CC mass fractions of Ru (solid line) and Ni (dash-dotted line) as a function of the isotopic composition of the CC endmember. The NC endmember is fixed as enstatite chondrites. The blue shaded area represents the mass fraction of CC-derived Mo in Nedagolla as given by its $\Delta^{95}$Mo value. The grey boxes represent the calculated mass fractions of CC-derived Ni and Ru if CI chondrites are used as the CC mixing endmember.

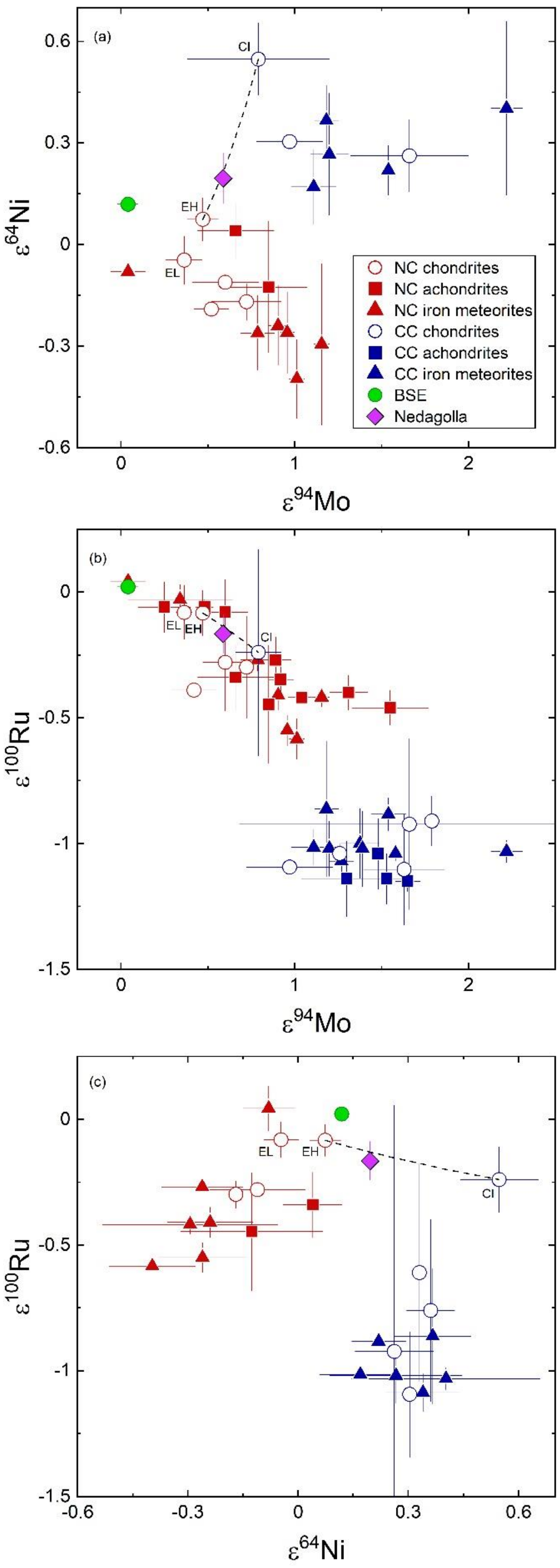

Fig. 7. Plots of (a) $\varepsilon^{64}$Ni vs. $\varepsilon^{94}$Mo, (b) $\varepsilon^{100}$Ru vs. $\varepsilon^{94}$Mo, and (c) $\varepsilon^{100}$Ru vs. $\varepsilon^{64}$Ni. Data for NC and CC meteorites from compilations in Burkhardt et al. (2019), Hopp et al. (2020), and Spitzer et al. (2020). Black dashed lines represent mixing lines between EH and CI chondrites, which in all three plots pass through the composition of Nedagolla. BSE – bulk silicate Earth.